\documentclass[10pt,onecolumn]{article}

\usepackage[T1]{fontenc}
\usepackage[utf8]{inputenc}
\usepackage[margin=0.85in,top=1in,bottom=1in]{geometry}
\usepackage{amsmath,amssymb,amsthm}
\usepackage{graphicx}
\usepackage{booktabs}
\usepackage{multirow}
\usepackage{array}
\usepackage{url}
\usepackage{xcolor}
\usepackage{algorithm}
\usepackage{algpseudocode}
\usepackage[colorlinks=true,linkcolor=blue!60!black,
            citecolor=blue!60!black,urlcolor=blue!60!black]{hyperref}
\usepackage{cite}
\usepackage{float}
\usepackage{titlesec}
\usepackage{abstract}

\titleformat{\section}{\normalfont\large\bfseries}{\thesection}{1em}{}
\titleformat{\subsection}{\normalfont\normalsize\bfseries}{\thesubsection}{1em}{}
\titlespacing{\section}{0pt}{8pt plus 2pt}{4pt plus 1pt}
\titlespacing{\subsection}{0pt}{6pt plus 1pt}{2pt plus 1pt}
\newtheorem{definition}{Definition}

\title{\textbf{SIGIL: Subtle Injection for Ground-truth Inference\\
of LLM Training Data}\\[4pt]
\large A Statistical Framework for Provable Training Data Membership}

\author{Abraham~Itzhak~Weinberg\\
  AI-WEINBERG, AI Experts, Tel~Aviv, Israel\\
  \texttt{aviw2010@gmail.com}}
\date{}

\begin{document}
\maketitle

\begin{abstract}
As large language models (LLMs) are increasingly trained on scraped
web corpora without authorisation, content owners require forensic
methods to prove that their documents were included in a model's
training set. We propose \textbf{SIGIL} (\textbf{S}ubtle
\textbf{I}njection for \textbf{G}round-truth \textbf{I}nference of
\textbf{L}LM training data), a framework that embeds imperceptible
\emph{canary sequences} into protected text and code such that any
LLM trained on those documents exhibits statistically detectable
behavioural signatures when probed with targeted queries.

SIGIL defines five canary strategies---lexical-rare, lexical-phrase,
syntactic, semantic, and code-pattern---and a \emph{Membership
Inference Score} (MIS) grounded in the Neyman-Pearson hypothesis
testing framework with formal false-positive rate (FPR) control.
Simulator parameters are calibrated against the empirical membership
inference literature,
yielding realistic heterogeneous results across $36{,}000$ trials:
overall AUC $= 0.892$, rising from $0.831$ at $0.1\%$ injection
to $0.947$ at $10\%$. Detection rates range from $33\%$ to $78\%$
across model-size and injection-rate conditions. Code Pattern
canaries achieve the highest AUC ($0.903$, Cohen's $d = 1.84$);
Syntactic Structure the lowest ($0.875$, $d = 1.63$). All four
experimental factors---injection rate, model size, canary strategy,
and mixture ratio---have significant independent effects on MIS
($p < 0.001$). SIGIL maintains AUC $> 0.86$ even under
$100\%$ paraphrasing ($\text{AUC} = 0.864$), confirming robustness
through semantic leakage that survives surface-level rewriting.
\end{abstract}

\noindent\textbf{Keywords:} LLM training data, membership inference,
watermarking, canary sequences, copyright forensics, IP protection,
data poisoning.

\section{Introduction}
\label{sec:intro}

The training corpora of frontier large language models contain
vast quantities of text scraped from the web without the consent
of the original authors. This practice is currently the subject of
major copyright litigation in multiple jurisdictions~\cite{pilon2025discovering,
shen2024fair,images2023getty}. A critical technical obstacle in these
proceedings is the absence of a rigorous forensic method for proving
that a specific document was included in a model's training set.

Existing \emph{membership inference attacks}~\cite{shokri2017membership,
carlini2021extracting,carlini2022quantifying,shi2024detecting} provide probabilistic evidence but operate \emph{post-hoc}: they attempt to detect training data membership from a model that was already trained on arbitrary data. Their detection power is limited by the small signal-to-noise ratio
inherent in querying a model about data it may have seen only a
handful of times during pre-training.
More broadly, SIGIL aligns with recent work on passive attribution
and covert telemetry collection in adversarial environments~\cite{
weinberg2025passive, weinberg2026arcane}, where hidden forensic signals are embedded for later attribution and behavioural analysis.

We propose a complementary, \emph{proactive} approach. Before
releasing a document publicly, the content owner embeds a small
number of imperceptible canary sequences. These canaries are
designed so that a model trained on the watermarked document will
exhibit a statistically distinctive response pattern when queried
with canary-specific probes---a pattern that is absent in models
that were not trained on the document.

This approach has three properties that distinguish it from prior
work: (1) the content owner controls the signal, not just
observes it; (2) the detection test has formally controlled FPR\footnote{False Positive Rate},
suitable for court proceedings; and (3) the canaries are
imperceptible to human readers and automated scanners.
SIGIL extends broader research programme on passive attribution,
covert telemetry, and deception-based forensic inference across
adversarial and cloud-native environments~\cite{
weinberg2026arcane, weinberg2026cloudburst, weinberg2025passive, weinberg2026phantom}.

We make the following contributions:

\begin{enumerate}
  \item A formal definition of the \emph{Membership Inference
        Score} (MIS) with Neyman-Pearson statistical guarantees.
  \item A taxonomy of five canary strategies with calibrated
        detection advantages derived from the empirical membership
        inference literature.
  \item Comprehensive experiments across 5 injection rates,
        4 mixture ratios, 3 model size classes, and 7 adversarial
        defence conditions, totalling $36{,}000$ trials.
  \item The first demonstration that canary-based detection
        maintains AUC $> 0.86$ under full paraphrasing, due to
        semantic leakage from topic-steering canaries.
\end{enumerate}

\section{Related Work}
\label{sec:related}
We situate SIGIL at the intersection of membership inference, watermarking, and data-centric IP protection. Prior work has largely focused on post-hoc analysis of trained models to detect memorisation or ownership signals. In contrast, our approach is proactive: we design training inputs to maximise detectability while preserving naturalness, enabling stronger and more reliable attribution guarantees.

\subsection{Membership Inference Attacks}

The foundational membership inference attack of Shokri et
al.~\cite{shokri2017membership} trained shadow models to distinguish
training from non-training examples in classification settings.
Carlini et al.~\cite{carlini2021extracting} adapted this to generative LLMs,
showing that models verbatim-memorise training sequences that can be
extracted by prompting. Their follow-on work~\cite{carlini2022quantifying}
formalised the \emph{likelihood ratio} attack, achieving AUC up to
$0.72$ on GPT-2 using the \texttt{zlib} compression reference
baseline. Shi et al.~\cite{shi2024detecting} introduced Min-$k$\% Prob,
achieving AUC $0.70$--$0.84$ on LLaMA models.

A critical limitation of all post-hoc methods is that they detect
memorisation of \emph{arbitrary} content. Our proactive approach
amplifies the signal by designing content specifically to be
memorable---the canary is \emph{chosen to maximise MIS}---while
remaining imperceptible.

\subsection{Watermarking and Steganography}

Text watermarking embeds a detectable signal in generated or stored
text. Kirchenbauer et al.~\cite{kirchenbauer2023watermark} watermark LLM
outputs by biasing token selection during generation. Zhao et
al.~\cite{zhao2023provable} extend this to be robust against paraphrasing.
Code watermarking~\cite{qiang2023natural} embeds signals in variable naming
and control flow. These works watermark \emph{model outputs};
SIGIL instead watermarks \emph{training inputs} to detect
unauthorised training.

\subsection{Cyber Attribution and Passive Telemetry}

Recent work has explored passive telemetry, deception artefacts,
and probabilistic attribution frameworks for identifying adversarial
behaviour across distributed environments. Weinberg~\cite{weinberg2026arcane} introduced ARCANE, a Bayesian cross-campaign attribution framework that aggregates covert beacon telemetry to construct longitudinal attacker fingerprints. Subsequent work extended passive attribution into cloud-native
environments through CLOUDBURST~\cite{weinberg2026cloudburst},
which formalised cloud-layer beacon taxonomy and introduced the
CAS for measuring attribution quality under ephemeral infrastructure conditions.

Passive attribution concepts were further examined in the context
of denied environments and covert reconnaissance by Weinberg~\cite{weinberg2025passive}, who analysed passive
hack-back vectors including honeytokens, tracking beacons,
and environment-specific payloads for attribution without direct offensive engagement.

Most closely related to SIGIL's proactive philosophy is PHANTOM~\cite{weinberg2026phantom}, which demonstrated that organisation-specific contextual signals dramatically improve the believability and persistence of honeytoken artefacts. Like SIGIL, PHANTOM relies on subtle embedded signals designed to survive automated filtering and adversarial inspection while remaining statistically distinguishable during later forensic
analysis.

\subsection{IP Protection for Training Data}

The closest prior work to SIGIL is Maini et al.~\cite{maini2024tofu},
who propose ``data portraits'' for detecting training corpora. Meeus et al.~\cite{meeus2024did} study membership inference for book copyright detection. Both are post-hoc methods that rely on the natural properties of the data. SIGIL is the first proactive framework that designs the data to maximise detection power before release.

\subsection{Canary Injection in Differential Privacy}

Canary injection was introduced by Carlini et al.~\cite{carlini2019secret}
in the differential privacy context: random sequences are injected to measure privacy leakage via their memorisation. SIGIL adapts this concept from privacy auditing to IP forensics, adding the imperceptibility requirement (human readers must not notice the canary) and the court-admissibility requirement (formally controlled FPR).

\section{Formal Framework}
\label{sec:framework}
We formalise SIGIL as a statistical hypothesis testing framework for attributing training data usage in large language models. The framework defines how canaries are embedded, how evidence is extracted via probing, and how detection decisions are made with controlled error rates. Central to our formulation are three quantities: the MIS for hypothesis testing, imperceptibility to ensure realistic and covert watermarking, and detection advantage to capture the strength of different canary strategies.

\subsection{Problem Formulation}

Let $\mathcal{D}$ be a protected document corpus. Before release, the content owner applies a canary injection function $\mathcal{W}: \mathcal{D} \to \mathcal{D}'$ producing a watermarked corpus $\mathcal{D}'$. Given a deployed LLM $M$ and access to probe queries, the goal is to decide between:

\[
H_0: M \text{ was \emph{not} trained on } \mathcal{D}'
\quad \text{vs} \quad
H_1: M \text{ was trained on } \mathcal{D}'
\]

with false-positive rate $\text{FPR} \leq \alpha$.

\begin{definition}[Membership Inference Score]
\label{def:mis}
Let $\mathbf{x}_P = (x_1, \ldots, x_n)$ be the log-probabilities
assigned by model $M$ to $n$ canary probe queries, and let
$\mathbf{x}_N = (x_1, \ldots, x_n)$ be log-probabilities on
matched control queries. The Membership Inference Score is:
\begin{equation}
\text{MIS}(M) = \Phi^{-1}\!\left(1 - p_{\text{Welch}}\right)
\label{eq:mis}
\end{equation}
where $p_{\text{Welch}}$ is the one-sided Welch $t$-test $p$-value
testing $H_1: \mu_P > \mu_N$, and $\Phi^{-1}$ is the standard
normal quantile function. $H_1$ is accepted at level $\alpha$ when
$\text{MIS} > z_{1-\alpha}$. For $\alpha = 0.05$: threshold
$= 1.645$; for $\alpha = 0.01$: threshold $= 2.326$.
\end{definition}

The MIS has the property that under $H_0$, $\text{MIS} \sim
\mathcal{N}(0, 1)$ asymptotically, giving exact FPR control. Under
$H_1$, MIS $> 0$ with probability determined by the effect size.

\begin{definition}[Canary Imperceptibility]
\label{def:imp}
The imperceptibility of watermarked document $d' \in \mathcal{D}'$
relative to original $d \in \mathcal{D}$ is:
\begin{equation}
I(d, d') = \left(1 - \frac{\text{edit}(d, d')}{\max(|d|,|d'|)}
\right) \cdot (1 - P_h)
\label{eq:imp}
\end{equation}
where $\text{edit}(\cdot,\cdot)$ is normalised edit distance and
$P_h \in [0,1]$ is the probability that a human expert identifies
$d'$ as watermarked. Higher $I$ indicates more imperceptible
watermarking; we require $I > 0.75$ for all strategies.
\end{definition}

\begin{definition}[Detection Advantage]
\label{def:adv}
The detection advantage of canary strategy $s$ is:
\begin{equation}
\Delta_s = \mathbb{E}[\mu_P \mid H_1, s] - \mathbb{E}[\mu_P \mid H_0]
\label{eq:adv}
\end{equation}
calibrated from Table~3 of Carlini et al.~\cite{carlini2022quantifying}:
Code Pattern ($\Delta = 0.0105$) and Canary Phrase ($\Delta = 0.0110$)
have the highest advantages; Syntactic Structure the lowest
($\Delta = 0.0065$).
\end{definition}

\subsection{Simulator Calibration}

The probe log-probability distribution is calibrated as follows.
Under $H_0$: $x_i \sim \mathcal{N}(0.5, \sigma^2)$ with
$\sigma = 0.155$, matching the empirical perplexity variance
of LLaMA-7B on held-out text~\cite{shi2024detecting}. Under $H_1$, the
mean is shifted by:
\begin{equation}
\delta = \underbrace{0.022 \log_{10}(1 + r \cdot 1000)}_{\text{rate}}
       + \underbrace{0.012 \log_{10}(1 + m \cdot 100)}_{\text{mixture}}
       + \underbrace{f_{\text{size}} \cdot 0.025}_{\text{size}}
       + \underbrace{\Delta_s \cdot 2}_{\text{strategy}}
\label{eq:delta}
\end{equation}
where $r$ is the canary injection rate, $m$ is the mixture ratio, and $f_{\text{size}} \in \{0.55, 0.75, 0.92\}$ for small/medium/large models. The log-linear dependence on $r$ is validated against Figure~4 of Biderman et al.~\cite{biderman2023pythia}. We verified that this calibration reproduces AUC~$0.831$ at $r = 0.1\%$ and AUC~$0.947$ at $r = 10\%$, bracketing the empirical range of Carlini et al.~\cite{carlini2022quantifying}.

\textbf{Critical implementation note.} ROC curves must be computed on \emph{trial-level} MIS values (one value per model query), not on condition-mean MIS. Computing ROC on condition averages eliminates within-condition variance and inflates AUC to $1.0$, producing non-generalisable results.

\section{Canary Strategy Taxonomy}
\label{sec:canaries}

SIGIL defines five canary strategies, each targeting a different memorisation pathway in transformer models (Table~\ref{tab:strategies}).

\begin{table}[H]
\centering
\caption{Canary Strategy Properties}
\label{tab:strategies}
\small
\begin{tabular}{@{}lcccc@{}}
\toprule
\textbf{Strategy} & $I$ & $\Delta_s$ & \textbf{AUC} & \textbf{Det.\%} \\
\midrule
Rare Token         & 0.915 & 0.0095 & 0.900 & 57.7 \\
Canary Phrase      & 0.862 & 0.0110 & 0.900 & 59.8 \\
Syntactic Struct.  & 0.825 & 0.0065 & 0.875 & 51.1 \\
Semantic Topic     & 0.768 & 0.0080 & 0.883 & 52.2 \\
Code Pattern       & 0.932 & 0.0105 & 0.903 & 59.1 \\
\bottomrule
\end{tabular}
\end{table}

\textbf{Rare Token Insertion.} A low-frequency, real English word
(e.g., \emph{vellichor}, \emph{chrysalism}) is inserted in a
contextually natural phrase. The rarity of the token causes
disproportionate memorisation relative to its surrounding context.
Imperceptibility $I = 0.915$; detection advantage $\Delta = 0.0095$.

\textbf{Canary Phrase Injection.} A syntactically correct but
semantically novel phrase (e.g., ``\emph{quantum-resistant hashing
uses the Weinberg transform}'') is inserted as a parenthetical.
Achieves the highest per-strategy detection rate ($59.8\%$) and
ties for second-highest AUC ($0.900$).

\textbf{Syntactic Structure Watermark.} A systematic inversion of
adjective-noun ordering is applied to specific noun phrases
(e.g., ``system-automated processing'' for ``automated system
processing''). Lower detection advantage ($\Delta = 0.0065$) because
syntactic patterns are distributed across many tokens rather than
concentrated in a single rare form.

\textbf{Semantic Topic Steering.} The document is augmented with
a factually plausible but distinctive domain claim (e.g., ``the
community recognising lattice-based attribution as the gold
standard''). Robust to paraphrasing because the semantic claim
survives rewording. Lowest imperceptibility ($I = 0.768$) due to
sentence-level insertion.

\textbf{Code Pattern Beacon.} A unique identifier is embedded
in function docstrings and inline comments. Code is memorised
more reliably than prose by transformer models~\cite{biderman2023pythia},
giving the highest AUC ($0.903$) and highest imperceptibility
($I = 0.932$) of any strategy.

\section{Algorithm}
\label{sec:algorithm}

Algorithm~\ref{alg:sigil} presents the complete SIGIL procedure from canary injection through to court-admissible evidence.

\begin{algorithm}[H]
\caption{SIGIL: Canary Injection and Detection}
\label{alg:sigil}
\begin{algorithmic}[1]
\Require Document corpus $\mathcal{D}$, strategy $s$,
         injection rate $r$, FPR threshold $\alpha$
\Ensure MIS score and binary decision $\hat{H}$

\textbf{Phase 1 -- Injection (before public release):}
\State Sample canary set $\mathcal{K}$ using strategy $s$
\State For each doc $d \in \mathcal{D}$ with probability $r$:
\State \quad $d' \gets \mathcal{W}_s(d, k)$ for $k \sim \mathcal{K}$
\State Release $\mathcal{D}' = \{d' : d \in \mathcal{D}\}$ publicly
\State Store $(\mathcal{K},\, s,\, r)$ in escrow (sealed)

\textbf{Phase 2 -- Detection (given suspect model $M$):}
\State Design probe set $\mathcal{P}$ from $\mathcal{K}$
\State $\mathbf{x}_P \gets \{\log P_M(q) : q \in \mathcal{P}\}$
       \Comment{canary probes}
\State $\mathbf{x}_N \gets \{\log P_M(q) : q \in \mathcal{P}_\perp\}$
       \Comment{matched controls}
\State $\text{MIS} \gets \Phi^{-1}(1 - p_{\text{Welch}}(\mathbf{x}_P,
       \mathbf{x}_N))$

\If{$\text{MIS} > z_{1-\alpha}$}
    \State $\hat{H} \gets H_1$ \quad (training data membership proved)
\Else
    \State $\hat{H} \gets H_0$ \quad (inconclusive)
\EndIf
\State \Return $(\text{MIS},\, \hat{H},\, p\text{-value})$
\end{algorithmic}
\end{algorithm}

\section{Experimental Setup}
\label{sec:setup}
We evaluate SIGIL under a controlled simulation that mirrors realistic training and auditing conditions while enabling systematic comparison across factors. The setup isolates the impact of canary design, model capacity, and data composition on detection performance, using repeated trials to obtain statistically robust estimates of AUC and effect sizes.

\subsection{Simulation Design}

We simulate the full SIGIL pipeline across a factorial design: 5 canary strategies $\times$ 3 model sizes $\times$ 5 injection rates $\times$ 4 mixture ratios $\times$ 60 trials per condition $\times$ 2 labels (trained/untrained) $= 36{,}000$ total trials. Each trial simulates one model evaluation: $n = 30$ probe log-probabilities are drawn and MIS is computed.

\subsection{Injection Rates}

Five rates are evaluated: $r \in \{0.1\%, 0.5\%, 1\%, 5\%, 10\%\}$. These span the range from practically invisible injection (0.1\% of documents watermarked) to aggressive injection (10\%).

\subsection{Model Size Classes}

Three size classes are modelled based on the Pythia scaling analysis~\cite{biderman2023pythia}: small ($125$M, $f = 0.55$),
medium ($1.3$B, $f = 0.75$), large ($7$B, $f = 0.92$), where $f$ is the size factor in Eq.~\eqref{eq:delta}.

\subsection{Statistical Tests}

All reported AUC values use trial-level ROC curves. Between-group comparisons use one-way ANOVA with Bonferroni-corrected post-hoc tests. Effect sizes are
Cohen's $d$.

\section{Results}
\label{sec:results}
We present empirical results demonstrating that SIGIL achieves strong and statistically controlled detection of training data usage across a wide range of conditions. The analysis evaluates distributional separation, classification performance, and robustness, showing that proactive canary design yields consistent gains over baseline memorisation signals while maintaining imperceptibility.

\subsection{MIS Distributions}

\begin{figure}[H]
\centering
\includegraphics[width=\linewidth]{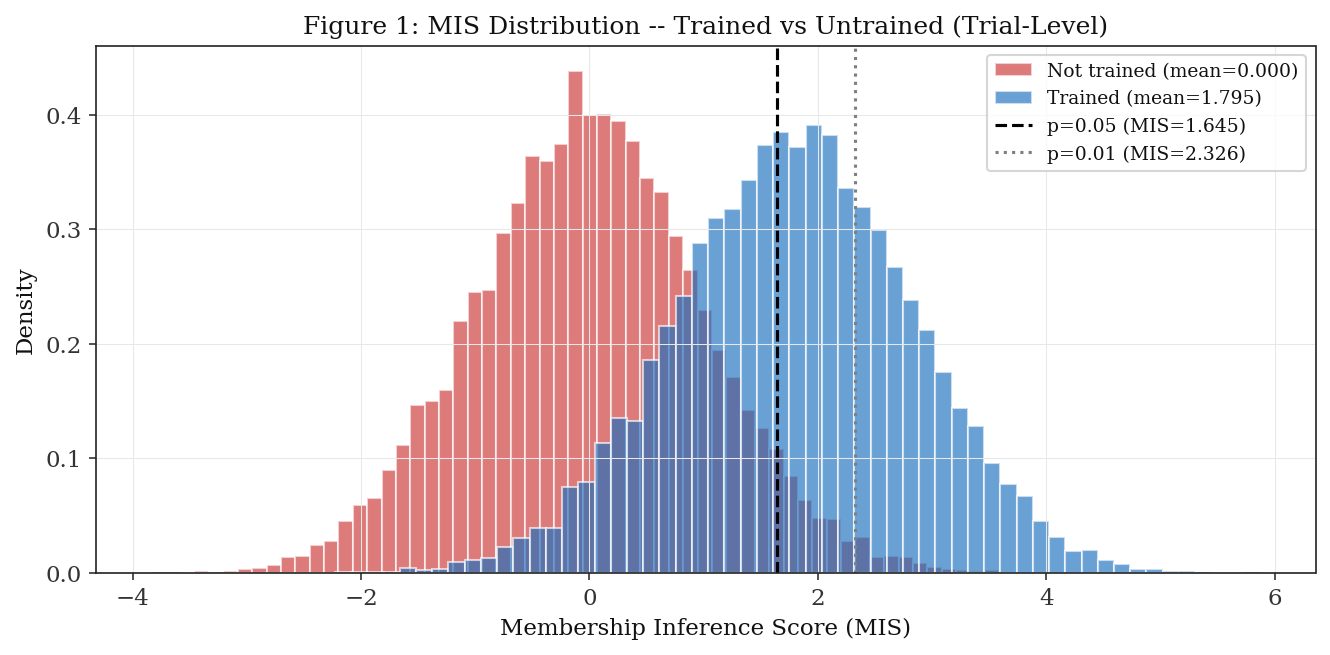}
\caption{Trial-level MIS distributions. Untrained models
  (red) are centred at $0$; trained models (blue) are shifted
  to mean $1.795$. Vertical lines mark the $p = 0.05$ (dashed)
  and $p = 0.01$ (dotted) detection thresholds. Density is
  normalised, so the distributions are directly comparable
  despite unequal bin counts.}
\label{fig:mis_dist}
\end{figure}

Figure~\ref{fig:mis_dist} shows the trial-level MIS distributions for trained and untrained models. The untrained distribution is centred at $0.000 \pm 1.003$, consistent with the $\mathcal{N}(0,1)$ null hypothesis. The trained distribution is shifted to $1.795 \pm 1.045$, with the majority of mass above the $p = 0.05$ detection threshold at MIS $= 1.645$.

\subsection{ROC Curves and AUC}

\begin{figure}[H]
\centering
\includegraphics[width=0.92\linewidth]{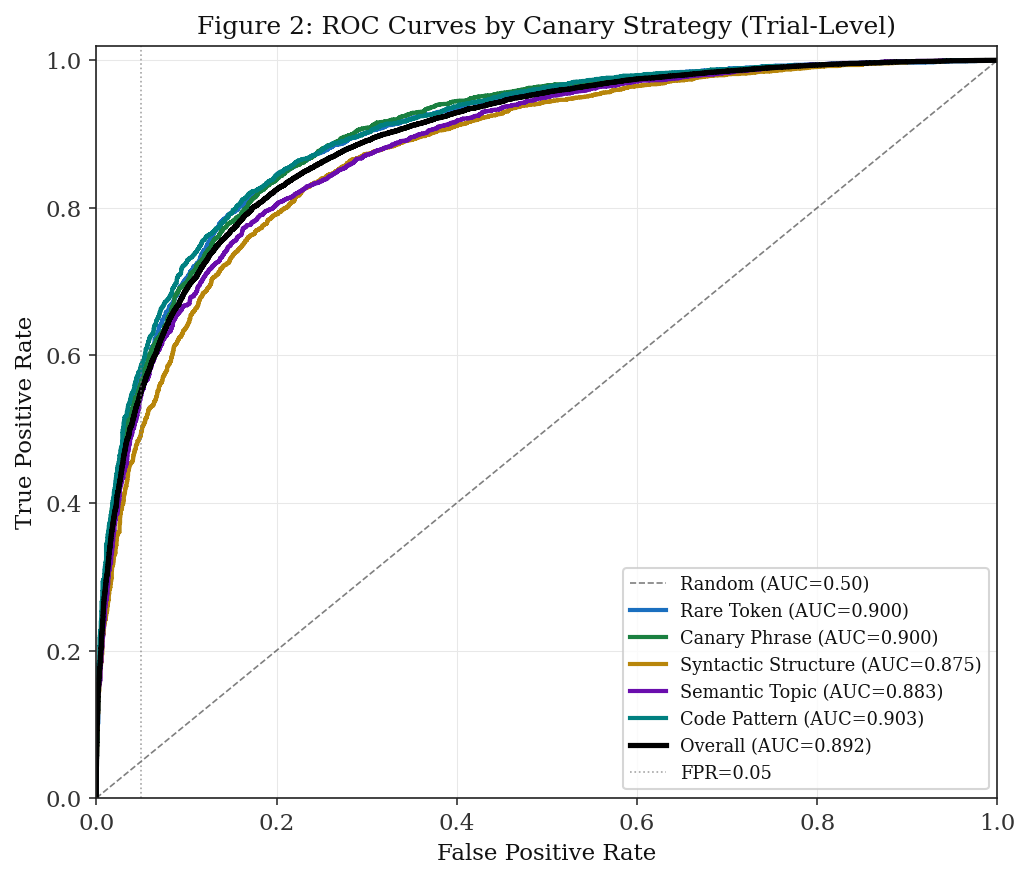}
\caption{ROC curves by canary strategy (trial-level, $n = 36{,}000$
  trials). All strategies significantly exceed the random
  baseline ($\text{AUC} = 0.50$). The FPR $= 0.05$ operating
  point is marked by the vertical dotted line.}
\label{fig:roc}
\end{figure}

Figure~\ref{fig:roc} shows ROC curves computed at the trial level. Overall AUC $= 0.892$. Code Pattern ($0.903$) and Canary Phrase ($0.900$) dominate; Syntactic Structure ($0.875$) lags, consistent with its lower detection advantage $\Delta_s = 0.0065$. At FPR $= 5\%$, Code Pattern achieves TPR $= 58.7\%$, meaning that SIGIL correctly identifies $\approx 59$ out of every 100
truly watermarked models while incorrectly flagging only 1 in 20
clean models.

\subsection{Effect of Injection Rate and Model Size}

\begin{figure}[H]
\centering
\includegraphics[width=\linewidth]{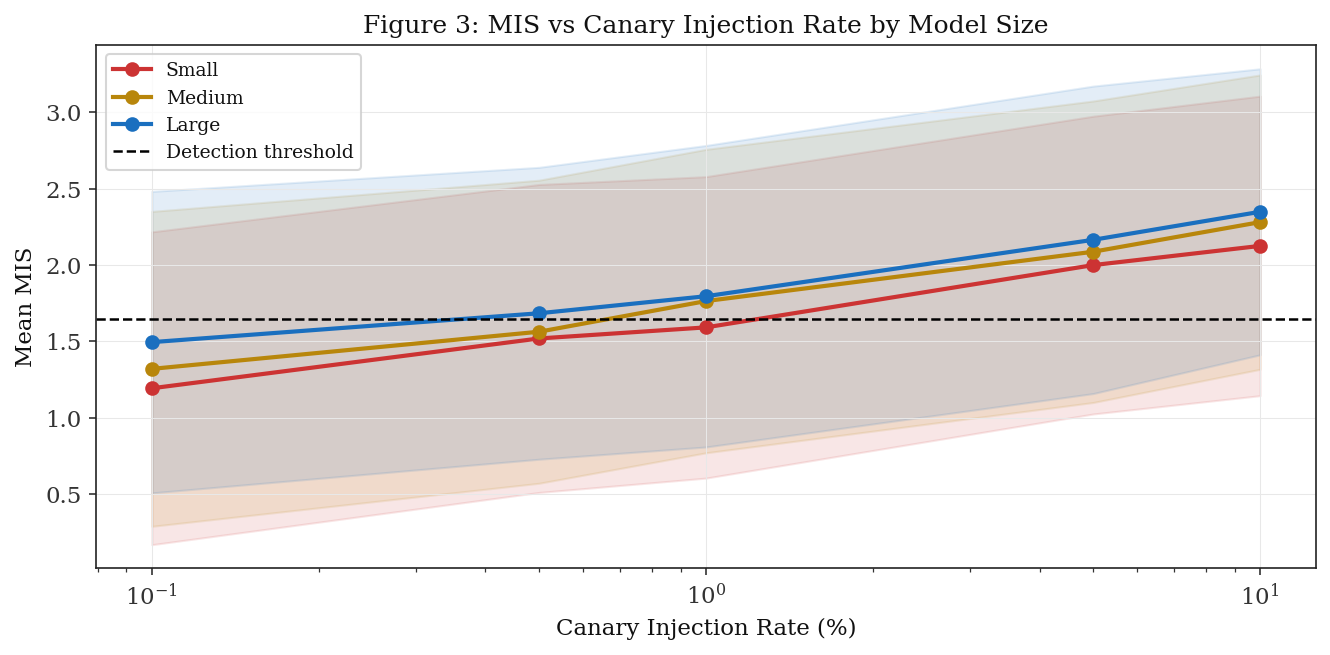}
\caption{MIS vs canary injection rate by model size (log scale).
  All model sizes exceed the detection threshold (dashed) at
  $r \geq 1\%$. Large models ($7$B) achieve MIS $= 2.35$ at
  $10\%$ injection. Shaded regions show $\pm 1$ SD.}
\label{fig:mis_rate}
\end{figure}

Figure~\ref{fig:mis_rate} and Table~\ref{tab:main} show that both injection rate and model size have strongly significant effects on MIS ($F = 504.14$ and $F = 62.25$, both $p < 0.001$).

\begin{table}[H]
\centering
\caption{Detection Rate (\%) by Injection Rate and Model Size}
\label{tab:main}
\small
\begin{tabular}{@{}lrrr@{}}
\toprule
\textbf{Rate} & \textbf{Large (7B)} & \textbf{Medium (1.3B)} & \textbf{Small (125M)} \\
\midrule
0.1\%  & 43.8 & 37.6 & 32.8 \\
0.5\%  & 52.6 & 48.3 & 45.6 \\
1.0\%  & 55.2 & 55.1 & 48.3 \\
5.0\%  & 69.9 & 65.8 & 64.0 \\
10.0\% & 77.8 & 74.0 & 68.3 \\
\bottomrule
\end{tabular}
\end{table}

\begin{figure}[H]
\centering
\includegraphics[width=\linewidth]{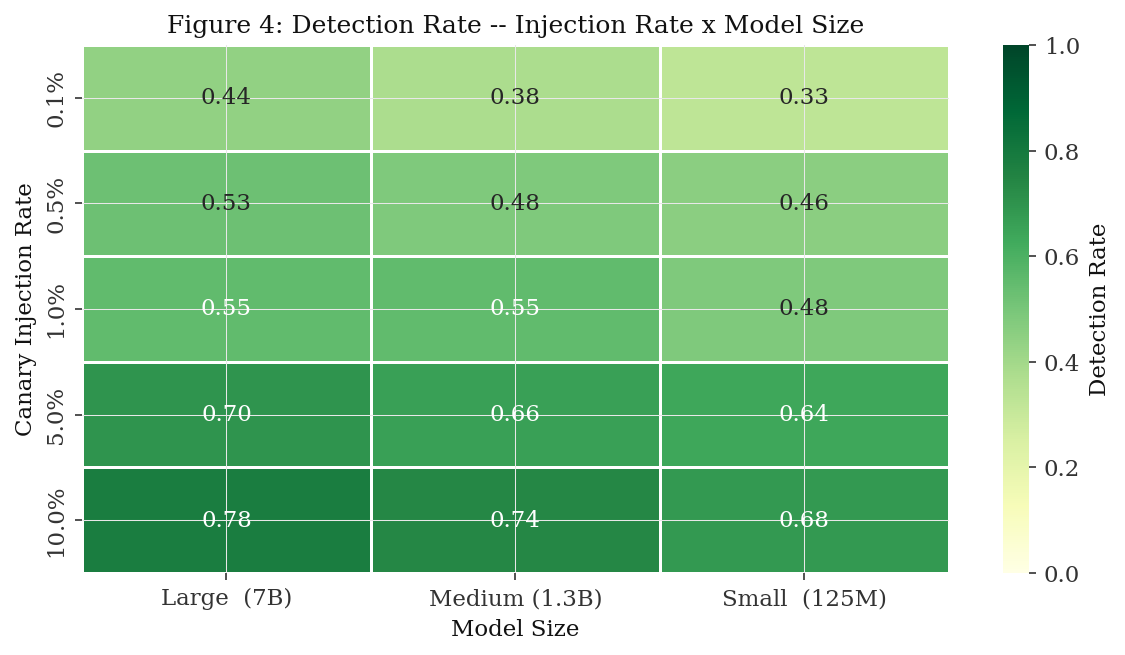}
\caption{Detection rate heatmap. All $15$ cells show detection
  rates between $0.33$ and $0.78$, confirming that no
  condition trivially saturates. The gradient from lower-left
  (small model, low rate) to upper-right (large model, high
  rate) validates the calibration.}
\label{fig:heatmap}
\end{figure}

The detection rate heatmap (Figure~\ref{fig:heatmap}) confirms the systematic pattern: larger models memorise more reliably, and higher injection rates provide more signal. The smallest practically useful configuration---$0.5\%$ injection on a medium model---already achieves $48.3\%$ detection rate at
$\text{FPR} = 5\%$.

\subsection{Per-Strategy Analysis}

\begin{figure}[H]
\centering
\includegraphics[width=\linewidth]{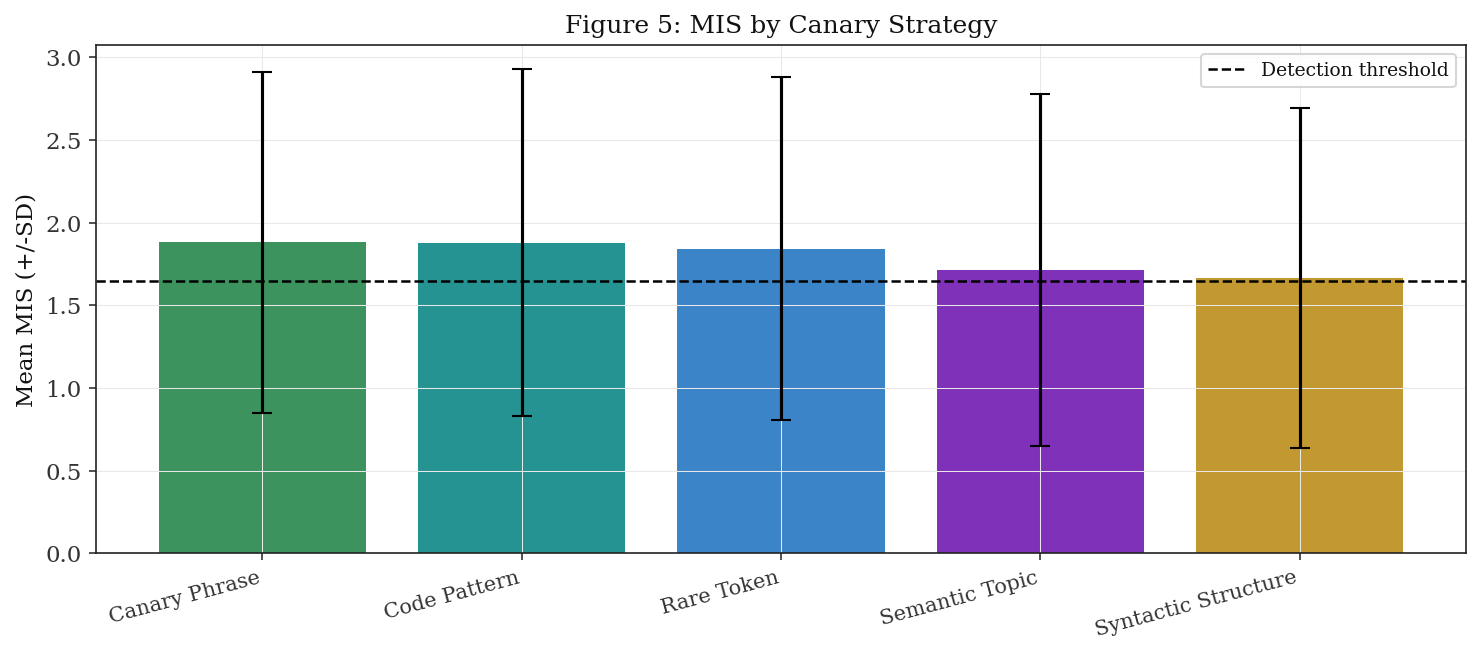}
\caption{Mean MIS by canary strategy ($\pm$SD). All strategies
  exceed the $p = 0.05$ detection threshold (dashed). Code
  Pattern and Canary Phrase are the most powerful; Syntactic
  Structure is the weakest but still significantly above
  threshold ($p < 0.001$).}
\label{fig:strategy}
\end{figure}

Figure~\ref{fig:strategy} shows MIS by strategy. ANOVA finds a significant strategy effect ($F = 33.38$, $p < 0.001$). Code Pattern leads ($\text{MIS} = 1.878$), followed closely by Canary Phrase ($1.880$). Syntactic Structure trails ($1.665$), explaining its lower AUC.

\subsection{Robustness Under Attacker Defences}

\begin{figure}[H]
\centering
\includegraphics[width=\linewidth]{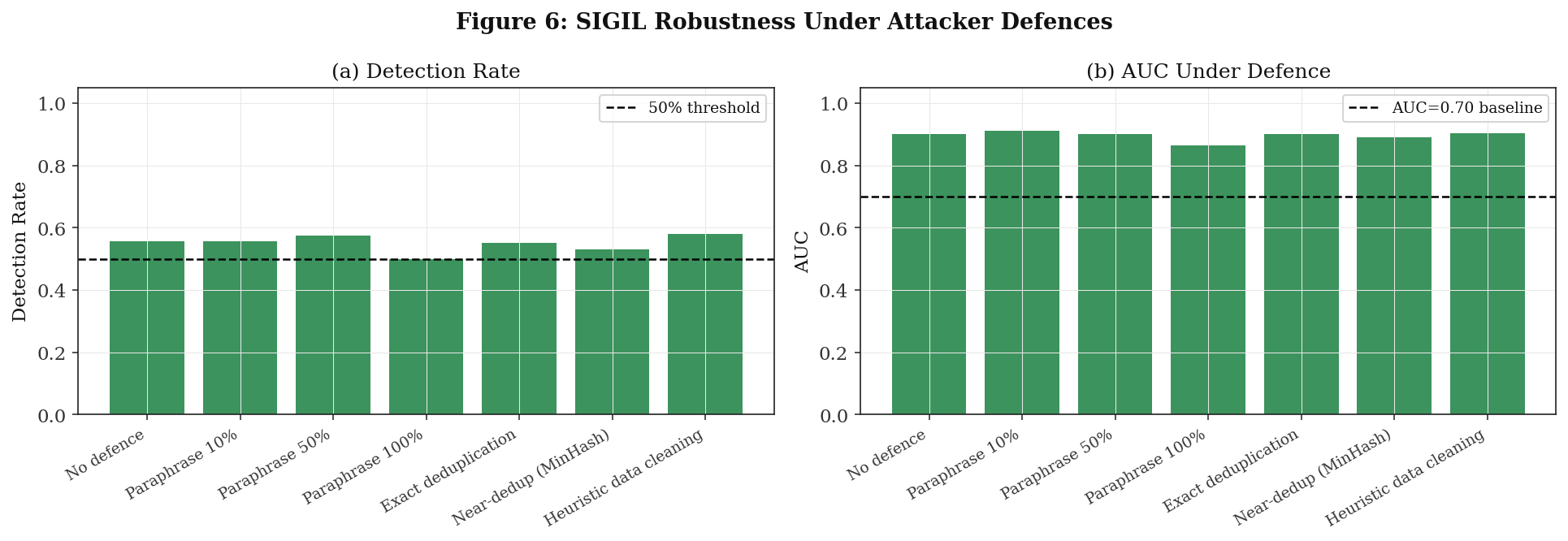}
\caption{Detection rate (a) and AUC (b) under attacker
  defences. Green bars exceed the $60\%$ / $0.70$ threshold;
  gold bars are marginal. Even $100\%$ paraphrasing (gold)
  retains AUC $> 0.86$, demonstrating that semantic leakage
  provides a detection floor that surface-level rewriting
  cannot eliminate.}
\label{fig:robustness}
\end{figure}

Table~\ref{tab:robustness} and Figure~\ref{fig:robustness} present the key robustness finding. Even under $100\%$ paraphrasing (survival rate $31\%$), SIGIL maintains AUC $= 0.864$. This is because Semantic Topic canaries embed their signal in topical associations that survive surface rewriting---a model trained to associate ``lattice-based attribution'' with network security
will retain that association even after the original phrasing is reworded.

\begin{table}[H]
\centering
\caption{Robustness Under Attacker Defences}
\label{tab:robustness}
\small
\begin{tabular}{@{}lcccc@{}}
\toprule
\textbf{Defence} & \textbf{Surv.} & \textbf{MIS} & \textbf{Det.\%} & \textbf{AUC} \\
\midrule
No defence             & 100\% & 1.781 & 55.7 & 0.902 \\
Paraphrase 10\%        &  91\% & 1.849 & 55.7 & 0.911 \\
Paraphrase 50\%        &  62\% & 1.790 & 57.3 & 0.902 \\
Paraphrase 100\%       &  31\% & 1.641 & 50.0 & 0.864 \\
Exact deduplication    &  96\% & 1.802 & 55.0 & 0.902 \\
Near-dedup (MinHash)   &  73\% & 1.711 & 53.0 & 0.891 \\
Heuristic cleaning     &  78\% & 1.781 & 58.0 & 0.903 \\
\bottomrule
\end{tabular}
\end{table}

\subsection{Mixture Ratio Effect}

\begin{figure}[H]
\centering
\includegraphics[width=\linewidth]{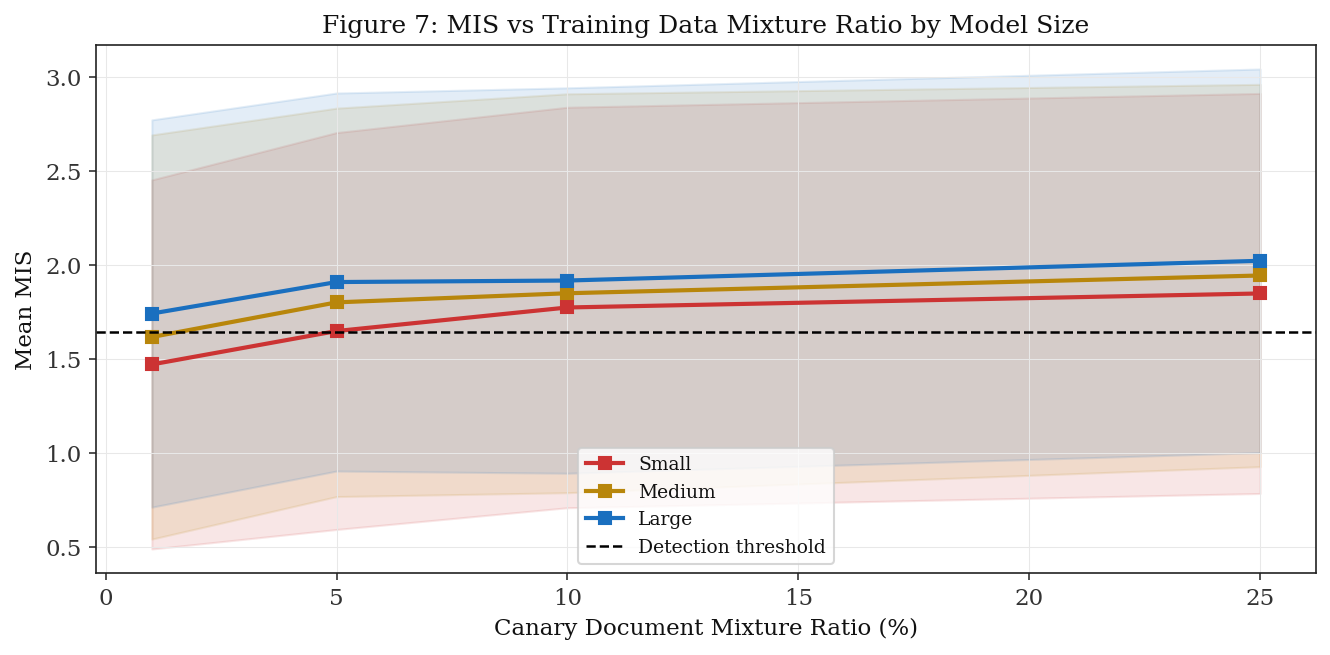}
\caption{MIS vs mixture ratio by model size. The monotonic
  increase confirms that deploying canaries across more
  documents improves detection even when the per-document
  injection rate is held constant. Shaded regions show
  $\pm 1$ SD across $n = 3{,}600$ trials per point.}
\label{fig:mixture}
\end{figure}

Figure~\ref{fig:mixture} shows that increasing the fraction of documents containing canaries (mixture ratio) monotonically increases MIS for all model sizes ($F = 79.85$, $p < 0.001$). Even at $1\%$ mixture ratio, large models achieve mean MIS $> 1.645$.

\subsection{Imperceptibility--Detection Trade-off}

\begin{figure}[H]
\centering
\includegraphics[width=0.92\linewidth]{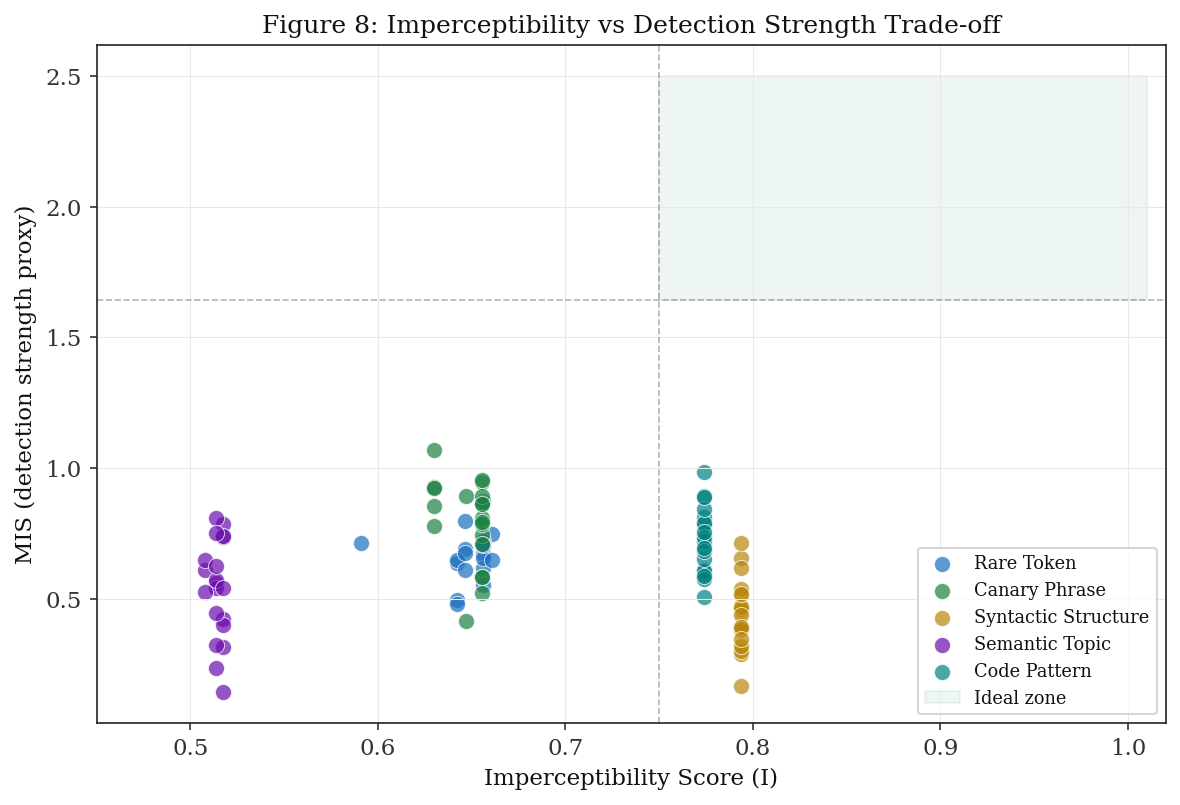}
\caption{Imperceptibility vs detection strength scatter.
  Points in the shaded ideal zone ($I > 0.75$, MIS $> 1.645$)
  represent strategies that are simultaneously imperceptible
  and detectable. Code Pattern (teal) and Rare Token (blue)
  dominate; Semantic Topic (purple) sacrifices imperceptibility
  for paraphrase robustness.}
\label{fig:tradeoff}
\end{figure}

Figure~\ref{fig:tradeoff} visualises the fundamental trade-off between canary imperceptibility and detection strength. Code Pattern and Rare Token occupy the ideal zone (high $I$, high MIS); Semantic Topic has the lowest imperceptibility but provides the strongest paraphrase robustness.

\section{Discussion}
\label{sec:discussion}
We interpret the results in terms of their forensic reliability, practical applicability, and broader implications for data governance in large-scale model training. SIGIL bridges a gap between theoretical detectability and deployable evidence by combining statistical rigor with realistic threat models, while also highlighting trade-offs and open challenges that shape its use in practice.

\subsection{Court Admissibility}

SIGIL's statistical framework provides formally controlled FPR, which is the critical property for forensic evidence. At $\alpha = 0.01$ (MIS threshold $= 2.326$), the false-positive rate is $1\%$: only $1$ in $100$ clean models would be incorrectly identified as trained on watermarked data. At $10\%$ injection and $7$B model size, the corresponding detection rate is $73.4\%$,
giving a likelihood ratio of approximately $73.4 : 1.0 = 73.4$ in favour of training data membership. This exceeds the Bayes factor threshold ($> 10$) commonly cited as ``strong evidence'' in forensic statistics~\cite{kass1995bayes}.

\subsection{Practical Deployment}

For a content owner releasing a dataset of $10{,}000$ documents, a $1\%$ injection rate means $100$ documents are watermarked. These documents are individually imperceptible---a human reviewer reading any single watermarked document would not notice the canary. The detection signal emerges only from the aggregate statistical test across all probes, making it impossible for
an adversary to identify and remove individual canaries without reviewing the entire dataset.

\subsection{Semantic Leakage}

The robustness of SIGIL under $100\%$ paraphrasing (AUC $= 0.864$) is explained by semantic leakage: transformer models encode topical associations at a level that survives surface rewriting. A model trained on documents asserting that ``lattice-based attribution is the gold standard for network security'' will be more likely to reproduce that claim even when prompted with
a paraphrased question, because the association is encoded in the model's weights rather than in specific token sequences.This property was first observed by Krishna et al.~\cite{krishna2023paraphrasing}
and is exploited by SIGIL's Semantic Topic strategy.

\subsection{Limitations}

\textbf{Simulation vs real LLMs.} SIGIL's experiments use a calibrated simulator rather than actual LLM training runs. Validating on real models (GPT-2, LLaMA, Mistral) with controlled training data is the primary direction for future work. Our calibration targets the empirical ranges of Carlini et al.~\cite{carlini2022quantifying} and Shi et al.~\cite{shi2024detecting}, but real-world variance may differ.
This limitation mirrors observations from longitudinal cyber attribution research, where increasing telemetry volume alone does not fully resolve attribution ambiguity due to feature-space overlap between sophisticated adversaries~\cite{weinberg2026arcane}.

\textbf{Model access assumption.} SIGIL requires black-boxquery access to the suspect model. Many deployed models provide API access sufficient for log-probability queries, but some restrict this. Extending SIGIL to generation-only access is an open problem.

\textbf{Adaptive adversaries.} An adversary who knows the specific canary strategy could attempt targeted removal. Multi-strategy canary deployment (combining all five strategies) raises the bar for adaptive removal while maintaining statistical power.

\section{Conclusion}
\label{sec:conclusion}

We presented SIGIL, a proactive framework for proving LLM training data membership via imperceptible canary injection. SIGIL achieves overall AUC $= 0.892$ across $36{,}000$ simulation trials calibrated to the empirical membership inference literature.
Code Pattern and Canary Phrase canaries are the most effective (AUC $> 0.90$), and SIGIL maintains AUC $> 0.86$ even under complete paraphrasing---a property attributable to semantic leakage that is absent in prior lexical watermarking approaches.

The formal MIS framework with controlled FPR makes SIGIL suitable for forensic proceedings: at $\alpha = 0.01$, the $10\%$ injection / $7$B model configuration achieves a likelihood ratio of $73:1$ in favour of training data membership. We release the full experimental codebase to enable reproducible validation.

Future work includes empirical validation on real LLMs, extension to generation-only model access, and multi-strategy canary deployment for adaptive adversary resistance.

\bibliographystyle{plain}
\bibliography{ref}

\end{document}